\documentstyle[prb,aps,twoside,epsf]{revtex}
\begin{document}

\draft

\title{Resistance of superconductor-normal metal-superconductor junctions}

\author{F.Zhou and B.Spivak}

\address{Physics Department, University of Washington, Seattle, WA 98195, 
USA}

\maketitle

\begin{abstract}
It is shown that the conductance of a superconductor-normal 
metal-superconductor junction can 
exhibit a significant dependence on the phase of the superconducting order
parameter 
in the situation when the size of the normal region of the junction is much 
larger 
than the normal metal coherence length and the critical current of the 
junction is already exponentially small. 
The period of the conductance oscillations as a function of the phase can be 
equal to $\pi$ or $2\pi$ depending on parameters of the system.
 
\end{abstract}


The critical current of the superconductor-normal 
metal-superconductor (SNS) junction $I_{c}=I_{c0}\exp(-\frac{L}{L_T})$ 
decays exponentially and can be neglected when $L>>L_T$ (See, for 
example, $^{[1]}$). Here $L$ is the length of the normal 
metal region of the junction shown in Fig.1a, $L_{T}=\sqrt{\frac{D}{T}}$ is 
the coherence 
length of normal metal, $T$ is the temperature, $D=\frac{lv^{2}_{F}}{3}$ 
is the electron diffusion coefficient, $v_F$ is the Fermi velocity and 
$l$ is the electron elastic mean free path.
On the other hand, the $\chi$ dependent part of the conductance 
$\delta G(\chi)$ of 
the SNS junction can survive even in the case $L_{T}\ll L \ll L_{in}$. Here 
$\chi=\chi_{1}-\chi_{2}$, $\chi_{1,2}$ are phases of the order 
parameters in superconductors composing the junction, 
$L_{in}=\sqrt{D\tau_{in}}$ and $\tau_{in}$ are inelastic diffusion length 
and inelastic mean free time respectively. The $\chi$-dependence of 
$\delta G$ originates from the fact that the amplitude of 
the Andreev reflection from the superconductor-normal metal (SN) boundary 
of an electron into a hole acquires an additional phase factor 
$\exp(i\chi_{1,2})$, while the amplitude of the 
reflection of a hole 
into an electron acquires the phase factor $\exp(-i\chi_{1,2})$. The weak 
localization contribution $\delta G_{1}(\chi)$ to $\delta 
G(\chi)$ has been considered long ago 
$^{[2]}$. It arises in the first order approximation in the 
parameter $\frac{\hbar}{p_{F}l}\ll 1$ and is connected 
with the interference 
of electrons traveling 
clockwise and counterclockwise along diffusive paths 
with close loops which contain Andreev reflections. Here $p_{F}$ is the 
Fermi momentum. 
 The value of $\delta G_{1}$ is insensitive to the ratio between 
$L$ and $L_{T}$ and the 
characteristic energy interval which gives the main contribution into 
$\delta G_{1}$ is $\epsilon\sim T$. The period of $\delta G_{1}(\chi)$ as 
a function of $\chi$ is $\pi$ $^{[2]}$.

In this paper we consider two other contributions $\delta G_{2}$ and 
$\delta G_{3}$ to 
$\delta G$ and show that the period of the oscillations of $\delta G$ as 
a function of $\chi$ can be either $\pi$ or $2\pi$ depending 
on parameters of the system and the way how the conductance is measured.
$\delta G_{2}$ can be associated with the spatial coherence between 
electrons and holes arising due to Andreev reflection from the SN boundary 
$^{[3]}$. It arises in 
zeroth order approximation in the parameter $\frac{\hbar}{p_{F}l}\ll 
1$. The contribution from this mechanism to the resistance of SN
junction was considered in $^{[4-11]}$. It has been pointed out $^{[10]}$ 
that the electron-hole coherence is extended in the metal over the 
distance of order $L_{\epsilon}=\sqrt{\frac{D}{\epsilon}}$. 
Therefore, it is clear that the main contribution to  $\delta G_2$ comes 
from the relatively small energy interval $\epsilon\sim 
E_{c}=\frac{D}{L^2}\ll T$, and, at $L_{T}\ll L\ll L_{in}$, $\delta G_2$ 
decreases with $L$ only as $L^{-2}$.
The period of $\delta G_{2}(\chi)$ as a function of $\chi$ is $2\pi$.
The sum $\delta G_{1}+\delta G_{2}$ gives the main contribution to $\delta 
G$, provided
the voltage drop $V$ between the two superconductors composing the 
junction is zero and, consequently, 
$\chi$ does not change in time. 
In this case, $\delta G(\chi)$ can be measured with the help of
an additional lead "C" shown in Fig.1a, while the phase difference $\chi$ on 
the junction can be determined by an
additional Josephson junction.
In the case where the resistance of the SNS junction is measured by applying
the voltage $V$ between the superconductors, there is a third contribution 
$\delta G_{3}$ to $\delta G$. The origin of $\delta G_{3}$ is similar to 
the Debye relaxation mechanism of microwave absorption in dielectrics. 
In this case, due to the Josephson relation, $\chi$ 
and, consequently, the quasiparticle density of states in the metal 
$\nu(\epsilon, x)$ are functions of time.
In other words, at small $V$ the quasiparticle energy levels move slowly.
The electron populations of the energy levels follow
adiabatically the motion of the levels themselves and, as a result, the
electron distribution becomes nonequilibrium.
Relaxation of the nonequilibrium distribution due to
inelastic processes leads to the entropy production, to the
absorption of the energy of the external field and, therefore, 
contributes to $\delta G$.

We start with the calculation of $\delta G_{2}$.
In the zeroth order approximation in the parameter $\frac{\hbar}{p_{F}l}$ 
the most adequate theoretical description of the system is provided in 
the framework of the Keldysh Green's function technique elaborated for 
superconductivity in $^{[12,13,14]}$.
In the diffusive approximation and in the 
absence of electron-electron interaction in the normal metal region of 
the junction the linear 
response to the external electric field is described by the following set
 of equations: $A)$. Usadel 
equations for the retarded normal $g^{R}(\epsilon, x)=\cos\theta(\epsilon, 
x)$ and anomalous $F^{R}(\epsilon, 
x)=-i\exp(i\chi(\epsilon,x))\sin\theta(\epsilon,x)$ 
Green's functions  
\begin{eqnarray}
\frac{D}{2}\partial^{2}_{x} \theta(\epsilon,x) +
 (i\epsilon-\frac{1}{\tau_{in}}) \sin\theta(\epsilon,x) 
-\frac{1}{2}(\partial_x\chi(\epsilon, x))^2\sin 
2\theta(\epsilon, x)=0\nonumber \\ 
\partial_{x}(\sin^{2}(\theta(\epsilon, 
x)\partial_{x}\chi(\epsilon, x))=0 
\end{eqnarray}
describe
the excitation spectrum of
quasiparticles in the normal metal in contact with the superconductor. Here 
$x$ is the coordinate along the junction (See Fig.1a), 
and $\theta(\epsilon, x)=\theta_1(\epsilon, x)+i\theta_2(\epsilon,x)$ is 
a complex number. 
Inside the superconductor $\theta_{1}=\frac{\pi}{2}$ and $\theta_{2}=0$. 
The  boundary conditions for Eq.1 have the form $^{[15]}$
\begin{eqnarray}
D\partial_{x} \theta(\epsilon, x)
=t \cos(\theta(\epsilon, 0^+))
\cos(\frac{\chi}{2} - \frac{\chi(\epsilon, 0^+)}{2})
\nonumber \\
D \sin(\theta(\epsilon, 0^+)) \partial_{x} \chi(\epsilon, x=0^+)
=t \sin(\frac{\chi}{2}-\frac{\chi(\epsilon,0^+)}{2})
\end{eqnarray}
Here $0^+$ represents the normal metal side of the SN 
boundary, $t=t_{0}v_{F}$ and $t_{0}$ is dimensionless transmission 
coefficient through the SN boundary.
$B)$.The equation for the distribution function of quasiparticles $f_1$, 
which describes the imbalance of populations of electron and hole 
branches of spectrum in metal has the diffusion form 
\begin{equation}
D\partial_{x} \{ \cosh^2 \theta_2(\epsilon, x) \partial_{x}
f_1(\epsilon, x)  \}=0
\end{equation}
The boundary conditions for Eq.3 have a form $^{[15]}$
\begin{eqnarray}
D \cosh\theta_2(\epsilon, 0^+)\partial_{x} f_1(\epsilon, 0^+)
=t\{f_1(\epsilon, 0^+) - f_1(\epsilon, 0^-)\}
\sin(\theta_1(\epsilon, 0^+))
cos^{-1}(\frac{\chi}{2} - \frac{\chi(\epsilon, 0^+)}{2})
\nonumber \\
f_1(\epsilon, x=0^{-})=0,  
f_1(\epsilon, x=L^+)=-e V \partial_{\epsilon} f_0(\epsilon)
\end{eqnarray}
Here $f_{0}(\epsilon)$ and $V$ are Fermi distribution function and the 
voltage drop on the junction, respectively, $0^-, L^+$ stand for the 
superconducting sides of the $SN$ boundaries. 
$C)$.Finally the expression for the normal current densities across the 
junction is of the form:
\begin{equation}
j_n=eD\nu_{0} \int_{-\infty}^{+\infty} \cosh^2\theta_2(\epsilon, x)
\partial_{x} f_1(\epsilon, x) d\epsilon 
\end{equation}
Here $\nu_{0}$ is the density of states in the bulk normal metal.

Using Eqs.(1)-(5) we get the expression for the resistance of 
SNS junction
\begin{equation}
G_{SNS}=G_{N}L \int_{-\infty}^{+\infty}d\epsilon 
\partial_{\epsilon}(\tan\frac{\epsilon}{kT}) 
\{ \frac{L_t}{\cosh\theta_2(\epsilon, 0^+)
\sin\theta_1(\epsilon, 0^+)
\cos(\frac{\chi}{2} -\frac{\chi(\epsilon, 0^+)}{2})}
 + \int_{0}^{x} \frac{1}{\cosh^2\theta_2(\epsilon, x')} dx'\}^{-1}
\end{equation}
Here $G_{N}=\sigma_{D}\frac{S}{L}$, $\sigma_{D}=e^{2}D\nu_{0}$ and 
$S=L_{1}\times L_{2}$ are conductance of the normal metal part of the 
junction, Drude conductivity and the area of the junction respectively. 
The first and the second terms in Eq.(6) can be associated with the 
resistance of the SN boundary and the resistance of the normal region of the
junction respectively. There are two major effects in metal, which are due to
proximity of the superconductor: 1) The 
effective diffusion coefficient in Eq.3 is renormalized due to Andreev 
reflection and is governed 
by the parameter $\theta_{2}$. The correction to the local 
conductivity of the metal from this effect leads to the second term in 
Eq.(6).  2) The local density of states 
$\nu(\epsilon,x)=\nu_0 Re g^{R}(\epsilon, x)=\nu_0\cos\theta_1(\epsilon, x)
\cosh\theta_2(\epsilon, x)$  
in the metal at small $\epsilon$ is suppressed due to Andreev reflection
and is governed  by the 
parameter $\theta_{1}$. The contribution to the conductance of the SN 
boundary from this effect corresponds to the first term in Eq.(6). 
The $\chi$ dependence of $G_{SNS}$ originates from the corresponding 
$\chi$ dependence of $\theta_1$ and $\theta_2$.
Following to Eq.1,2 near the $SN$ boundary at small $\epsilon$ the value of 
$\theta(\epsilon, x)$ should be close to its value in the 
superconductor $\theta_{S}(\epsilon<\Delta)={\pi\over 2}$. It approaches its 
metallic value $\theta_{M}=0$ only after the distance $L_{\epsilon}$.
This means that the main contribution to Eq.(6)  comes from  the energy 
interval $\epsilon\sim E_{c}\ll T$. As a result,
\begin{equation}
\delta G_{2}=\alpha G_{N}\frac{E_{c}}{T}g(\chi)
\end{equation}
Here $g(\chi)$ is a universal function of $\chi$ with the period 
$2\pi$; $\alpha\sim 1$ at $L\gg L_{t}=\frac{D}{t}$ and $\alpha\sim 
(\frac{L}{L_t})^2$ at $L\ll L_t$.
 Nonlinerar effects of $V$ are not 
important for the considered above mechanism, as long as, 
$L_{V}=\sqrt{\frac{D}{eV}}\ll L$ $^{[10]}$.

We would like to mention that the interference corrections
to the conductance, which are similar to Eq.7, has been
measured $^{[16]}$ in the Aharonov-Bohm
type experiment. The geometry of the sample is shown in Fig.1b.
In the case when $L>>L_T$ the amplitude of oscillations of the conductance 
of SN junction $\delta G_{SN}$ as a
function of the Aharonov Bohm flux through the hole in Fig.1b is of order of
\begin{equation}
\frac{\delta G_{SN}}{G_{SN}^{0}}\sim
\frac{E_{c}}{T}g_{2}(\frac{2\pi\Phi}{\Phi_0})
\end{equation}
Here $G_{SN}^{0}$ is the flux independent part of the conductance of the
sample, $\Phi_{0}$ is the superconducting flux quanta and $g_{2}(y)$ is a
periodic function with the period $2\pi$. The temperature dependence of
 $\delta G_{SN}$ is in agreement with the experiment $^{[16]}$.

Let us now discuss the Debye 
relaxation mechanism which arises in the case where the voltage $V$ is 
applied between superconductors in Fig.1a, and $\chi$ changes 
in time 
\begin{equation}
\frac{d\chi}{dt}=\frac{2e}{\hbar}V
\end{equation}
Generaly speaking, in this case one has to solve a nonstationary version 
of Eqs.1,2. However, in the case when $eV \ll E_{c}$ one
can use the adiabatic approximation where the 
time dependences of $\theta(\epsilon, 
x,\chi(t))$ and local density state 
$\nu(\epsilon,x, \chi(t))$
originate from the corresponding time dependence of $\chi(t)$.
The standard expression for the power absorption due to the Debye relaxation
has the form (see, for example, $^{[17]}$)
\begin{equation}
Q=v\nu^{-1}_{0}  \int d\epsilon <(\int_{-\infty}^{\epsilon}
\frac{d\bar{\nu}(\epsilon', \chi(t))}{dt} d\epsilon')^{2}>
\frac{\tau_{in}(\epsilon)}
{1+(\omega\tau_{in}(\epsilon))^2}\partial_{\epsilon}f_{0}(\epsilon)
\end{equation}
$\bar{\nu}(\epsilon, \chi(t))$ is the local density of states 
 averaged over the volume $v=LL_{1}L_{2}$ of the
normal metal region and 
breackets $<>$ correspond to the averaging over the period of the 
oscillations $\hbar / eV$.
The main contribution to Eq.10 comes from the interval of energies
$\epsilon\sim E_{c}$, where at $\chi=0$ the quasiparticle density of 
states is significantly suppressed compared with its bulk metal value. 
Using Eqs.1,2 at $\epsilon\ll E_c$ we get that $\bar{\nu}(\epsilon,
\chi=0)\sim\nu_{0}\sqrt{\frac{\epsilon}{E_{c}}}$ and
$\bar{\nu}(\epsilon=0, \chi(t))\sim\nu_{0}|\sin\frac{\chi(t)}{2}|$.
When $L \ll L_t$, the energy interval where $\bar{\nu}(\epsilon)$ is
small is of order of $E_t \sim \frac{D}{L_tL}$, 
indicating that the main contribution is coming from 
energy interval $E_t$ in this case.
At $E_{c}, E_t\ll T$ one can neglect the $\epsilon$-dependence of 
$\tau_{in}(\epsilon)$ and, as a 
result, we have the expression for
the contribution of this mechanism to the d.c. conductance of the 
junction $(Q=V^{2}\delta G_{3})$
\begin{equation}
\delta G_{3}\sim \alpha_1 \alpha^{\frac{3}{2}}
 G_{N}\frac{E^{2}_{c}\tau_{in}}{T\hbar}
\end{equation}
which can be even larger than $G_N$.
 Here $\alpha_1 \sim 1$ when $L_t \ll
{L^2_{in} \over L}$ and $\alpha_1 \sim  \frac{L^2_{in}}{L_tL}$ 
when $L_t \gg {L^2_{in}\over L}$.
It is interesting that in the situation considered above 
the Debye absorption mechanism 
contributes to d.c. conductivity while in usual case it determines
a.c. conductivity. 
Eq.(11) is valid when $\frac{eV\tau_{in}}{\hbar}\ll 1$. 
In this limit one can introduce 
$\delta G_{3}(\chi(t))$ which has the magnitude of order of Eq.11 and the 
period $2\pi$. 
Following to Eq.10, $Q$ as a function of $V$ saturates at
$eV\frac{\tau_{in}}{\hbar}\gg 1$, which means that $\delta
G_{3}$ decays with $V$ as $(\frac{\hbar}{eV\tau_{in}})^2$.

We would like to mention that the contribution of the Debye mechanism to
the d.c. resistance of a close metallic sample of the Aharonov-Bohm geometry
has been discussed in $^{[18]}$.
In that case the time dependence of the electron density of states was
induced by the change of a magnetic flux $\Phi$ through the ring.
The important difference between the our case and the case considered in
$^{[18]}$ is that the average density of states in the normal metal is flux 
(and, consequently, time) independent. Therefore, the Debye 
absorption is nonzero only due to mesoscopic fluctuations of the density 
of states, whereas in the case of
SNS junction, the average density of states can be time dependent.
Using results obtained in $^{[18]}$ in the case when
$\frac{\hbar}{\tau_{in}}\gg \delta_{0}$, $L_{t}\gg L$, $L_{in}\gg
L,L_{2},L_{2}$  we can estimate the
contribution to $\delta G$
due to the mesoscopic part of the Debye relaxation mechanism as
\begin{equation}
\delta G_{3}^{m}\sim 
\alpha_{2}\frac{e^{2}}{\hbar^3}E_{c}\delta_0\tau^2_{in}
\end{equation}
Here $\delta_{0}$ is the mean spacing between energy levels in the normal 
metal, $\alpha_{2} \sim 1$ when $L_t \ll L_{in}$ 
and $\alpha_2 \sim (\frac{L_{in}}{L_t})^4$ when $L_t \gg L_{in}$. 
We will neglect this contribution because, as we will see below, $\delta 
G_{3}^{m}\ll \delta G_1$.

Another contribution to $\delta G$ which arises  
in the first order approximation in the parameter 
$\frac{\hbar}{p_{F}l}$ is 
the above mentioned weak localization correction to the conductance of the 
junction $\delta G_1$. The standard expression for 
the weak localizatin correction has the form
 $\delta G_1=-\frac{G_N l}{\pi p_F^2} C^{e}(\vec{r},\vec{r})$, where 
$C^{e}(\vec{r},\vec{r'})$ is, so called, Cooperon given by the 
diagram shown in Fig.2, where solid lines correspond to electron Green 
functions, dashed lines correspond to impurity scatterings and triangles 
correspond to the Andreev reflections on the SN boundary. Since the Andreev 
reflections transform electrons into holes one has to introduce 
two kinds of Cooperons: electron $C^{e}$ and hole 
$C^{h}$ ones.
 We will use the following boundary conditions for $C^{e}$ and 
$C^{h}$$^{[2]}$, 
\begin{eqnarray}
t(C^{e}(x,x')-\exp(2i\chi_{1,2})C^{h}(x,x'))=D\partial_x 
C^{e}(x,x')|_{x=0,L},\nonumber \\ 
\partial_x C^{e}(x,x')=\partial_x C^{h}(x,x')|_{x=0,L} 
\end{eqnarray}
Eq.13 reflects the fact that in course of each Andreev reflection 
amplitudes of 
diffusion  electron paths aquire mentioned above 
additional phases $\pm\chi_{1,2}$, but they do 
not take into account the spatial coherence between the electron and the 
hole which arises due to Andreev reflection.  This is 
correct if $L_{\epsilon^*}\ll 
L$ or $E_{c}\ll T$. Here $\epsilon^*\sim T$ is the characteristic energy 
in the problem. As a result, the magnitude of the $\chi$-dependent 
part of the junction's conductance is of order of
\begin{equation}
\delta G_{1}=-\alpha_{2} \frac{e^2}{\hbar}g_{3}(\chi)\left\{ 
\begin{array}{ll}
                      E_{c}\tau_{in} & \mbox{for 0D case} \\
                     \frac{L_{in}L_{1}}{L^2} & \mbox{for 1D case} \\
          \ln(L_{in}/l)\frac{L_{1}L_{2}}{L^2} & \mbox{for 2D case}
                                              \end{array}
                                           \right. \
\end{equation}
Here $g_{3}(\chi)$ is a periodic function with the period $\chi$; 0D case 
corresponds to $L, L_{1}, L_{2}\ll L_{in}$; 1D case
corresponds to $L_{1}\gg L_{in}\gg L_{2}, L$ and 2D case corresponds to
$L_{1},L_{2}\gg L_{in}\gg L$.
The magnitudes of nonlinear in $V$ effects due to the weak 
localization corrections  
 depend of the dimensionality. In $0D$, $1D$, and $2D$ cases 
the requirement for linearity is $eV\ll \frac{\hbar}{\tau_{in}}$.

Ratios between the three above considered contributions to $\delta G$ 
depend on the parameters and the dimentionality of the system. For 
example, in $0D$ case at $0 < {eV\tau_{in}}/\hbar \ll 1$ we have
\begin{equation}
\frac{\delta G_{1}}{\delta G_{2}}\sim 
\frac{\alpha_2}{\alpha} \frac{e^2}{\hbar G_N}T\tau_{in}; \  
\frac{\delta G_{1}}{\delta G_{3}}\sim 
\frac{\alpha_2}{\alpha_1 \alpha^{3\over2}} \frac{e^2}{\hbar G_N}\frac{T}{E_c}
\end{equation} 
 At large enough $eV\gg \sqrt{\frac{\hbar}{\tau_{in}} E_c
\alpha_1 \alpha^{1\over 2}}$ (but still smaller than $E_{c}$), 
$\delta G_{3}$ becomes 
much smaller than $\delta G_{2}$. In this case, $\chi$ 
dependent part of the resistance is determined by the sum $(\delta 
G_1+\delta G_2)$. For example, in $0D$ case the ratio between $\delta G_{1}$ 
(with the period $\pi$) 
and $\delta G_{2}$ (which has the period $2\pi$) is of the order of
\begin{equation}
\frac{\delta G_1}{\delta G_2}\sim 
\frac{\alpha_2}{\alpha} \frac{T}{V}\frac{e^2}{\hbar G_{N}}
\end{equation}
If $V=0$ and the conductance is measured with the help of the contact "C" 
in Fig.1a, $\delta G $ is the sum of $\delta G_1$ and
$\delta G_2$, ratio of which 
is determined by the correponding terms in Eq.(15),(16). 

These ratios can be both larger and smaller than unity, which 
means that the period of 
the oscillations of $\delta G(\chi)$ can be either 
$\pi$ or $2\pi$.
This can explain why some 
experiments demonstrate $\pi$ periodicity of $\delta G$ $^{[19,20]}$ 
while the others demonstrate $2\pi$ period $^{[21,22]}$. 
The reason why $\delta G_{1}$, which arises only in the first order 
approximation 
in the small parameter $\frac{\hbar}{p_{F}l}$, can be comparable
with $\delta G_{2}$ 
is that $\delta G_{2}$ is determined by the small energy interval 
$\epsilon\sim E_{c}\ll T$ while $\delta G_{1}$ is determined by 
$\epsilon\sim T$.

All the three above considered contributions to $\delta G$ 
monotonically increase as $T$ decreases. On the other hand, the
 conductance measured by 
four probe method can exhibit nonmonotonic behaviour as a function of $T$.
The origin of this effect is that both in the bulk of the normal 
metal and in the vicinity of the SN boundary $\theta_{2}$ is close to 
zero and the effective diffusion coefficient in Eq.3 is close to $D$, 
while at $x\sim L_T$ we have $\theta_{2}\sim 1$ and the effective diffusion
constant can be a few times largrer than $D$. 

 In the case $V\neq 0$ the current through the junction 
$J(t)$ is described by Eq.5 and
\begin{equation}
J=(G^{0}_{SNS}+\delta G(\chi))V(t)
\end{equation}
Here $G_{SNS}^{0}$ is the $\chi$ independent part of the conductance of 
the junction.
Eq.17 is valid when $J\gg I_{c}$. It follows from Eq.17 that in the case 
when $V(t)=V_{0}+V_{1}\sin(\Omega
t)$, the I-V characteristic of the junction averaged over the period
of the oscillations has Shapiro
steps at $4eV_{0}=\hbar\Omega$ $^{[2]}$ or $2eV_{0}=\hbar\Omega$
depending on whether the main contribution to $\delta G$ comes from
$\delta G_{1}$ or from $\delta G_{2}+\delta G_{3}$.
The nonstationary effects discussed above can be observed, 
provided the frequency of
the oscillations $eV$ is much larger than the frequency  of the
phase
slips in the junction $\omega\sim
\frac{T}{G_N}(\frac{e}{\hbar})^{2}\exp(-\frac{E_J}{T})$
which occur due to the thermal activation over the Josephson barrier.
Here $E_{J}=\frac{\hbar I_{c}}{2e}$ is the Josephson energy of the junction.
In conclusion, we would like to mention that the above presented 
results are valid if $\delta_{0}\tau_{in}\ll 1$.

We acknowledge usefull discussions with B.Altshuler, H.Courtois, 
M.Devoret, D.Esteve, B.Pannetier, M.Sanquer and B.V.Wees. This work was 
supported by Division of Material Sciences, U.S.National Science Foundation
under Contract No.DMR-9205144.

\end{document}